# The cross-frequency mediation mechanism of intracortical information transactions

RD Pascual-Marqui[1,2], P Faber[1], S Ikeda[2], R Ishii[3], T Kinoshita[2], Y Kitaura[2], K Kochi[1], P Milz[1], K Nishida[2], M Yoshimura[2]

[1]The KEY Institute for Brain-Mind Research, University of Zurich, Switzerland; [2]Department of Neuropsychiatry, Kansai Medical University, Osaka, Japan; [3]Department of Psychiatry, Osaka University Graduate School of Medicine, Osaka, Japan

Corresponding author: RD Pascual-Marqui
pascualmarqui@key.uzh.ch ; www.uzh.ch/keyinst/loreta.htm
scholar.google.com/citations?user=pascualmarqui

## 1. Abstract

In a seminal paper by von Stein and Sarnthein (2000), it was hypothesized that "bottom-up" information processing of "content" elicits local, high frequency (beta-gamma) oscillations, whereas "top-down" processing is "contextual", characterized by large scale integration spanning distant cortical regions, and implemented by slower frequency (theta-alpha) oscillations. This corresponds to a mechanism of cortical information transactions, where *synchronization of beta-gamma oscillations between distant cortical regions is **mediated** by widespread theta-alpha oscillations*. It is the aim of this paper to express this hypothesis quantitatively, in terms of a model that will allow testing this type of information transaction mechanism. The basic methodology used here corresponds to statistical mediation analysis, originally developed by (Baron and Kenny 1986). We generalize the classical mediator model to the case of multivariate complex-valued data, consisting of the discrete Fourier transform coefficients of signals of electric neuronal activity, at different frequencies, and at different cortical locations. The "mediation effect" is quantified here in a novel way, as the product of "dual frequency RV-coupling coefficients", that were introduced in (Pascual-Marqui et al 2016, http://arxiv.org/abs/1603.05343). Relevant statistical procedures are presented for testing the cross-frequency mediation mechanism in general, and in particular for testing the von Stein & Sarnthein hypothesis.

## 2. Introduction

In a seminal paper by von Stein and Sarnthein (von Stein and Sarnthein, 2000), it was hypothesized that "bottom-up" information processing of "content" elicits local, high frequency (beta – gamma) oscillations, whereas "top-down" processing is "contextual", characterized by large scale integration spanning distant cortical regions, and implemented by slower frequency (theta – alpha) oscillations.

Quoting from (von Stein and Sarnthein, 2000): "In particular, large-scale low frequency interactions might allow an integration with the different local, fast gamma processes, with which sensory information from the periphery seems to be propagated 'bottom-up'."

These ideas have been extensively reviewed and discussed in (Engel et al., 2001). Experimental support can be found in e.g. (Buschman and Miller, 2007). Buzsaki and Wang (Buzsaki and Wang, 2012) describe this mechanism as "Brain-Wide Synchronization of Gamma Oscillations by Slower Rhythms".

Informally, the von Stein & Sarnthein hypothesis can be expressed as:
*"Synchronization of beta-gamma oscillations between distant cortical regions is <u>mediated</u> by theta-alpha oscillations."*

It is the aim of this paper to express this hypothesis quantitatively, in terms of a model that will allow testing this type of information transaction mechanism.

The basic methodology used here corresponds to statistical mediation analysis, originally developed by (Baron and Kenny, 1986). More recent textbook expositions of this methodology are available in (MacKinnon, 2008;Hayes, 2013).

In this work, we generalize the classical mediator model to the case of multivariate complex-valued data, consisting of the discrete Fourier transform coefficients of signals of electric neuronal activity, at different frequencies, and at different cortical locations. The "mediation effect" is quantified here in a novel way, as the product of "dual frequency RV-coupling coefficients", that were introduced in (Pascual-Marqui et al., 2016). Relevant statistical procedures are presented for testing the cross-frequency mediation mechanism in general, and in particular for testing the von Stein & Sarnthein hypothesis.

## 3. The simple univariate real-valued mediation model

This type of model, in its simplest form, consists of three variables: the independent variable "x", the mediator variable "z", and the dependent variable "y". Assuming that all variables are centered (i.e. have zero mean), they satisfy the following relations:

**Eq. 1** $\quad y = bz + cx + \varepsilon_y$

**Eq. 2** $\quad z = ax + \varepsilon_z$

where "a", "b", and "c" are regression coefficients, and $\varepsilon_y$ and $\varepsilon_z$ are noise.

The graphical representation of this model is shown in Figure 1, where the influence of "x" on "y" has two contributions: a direct influence (red arrow), and a mediated influence (the two blue arrows).

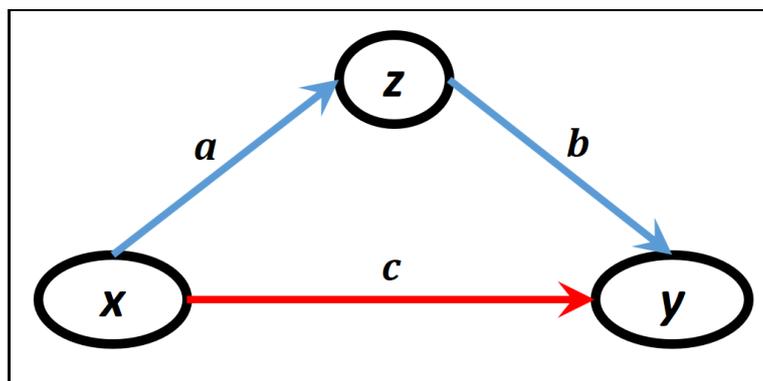

Figure 1: The classical mediation model, where "x" is the independent variable, "z" is the mediator variable, "y" is the dependent variable, and "a", "b", "c" are regression coefficients (see Eq. 1 and Eq. 2). The blue arrows correspond to the "mediation path", and the red arrow corresponds to the "direct path".

The mediation effect is defined and quantified as:

**Eq. 3** $\quad E_1(x \to z \to y) = ab$

If this value is zero, then there is no mediation effect. For this to happen, one or both regression coefficients must be zero.

Thus, an alternative way to quantify the mediation effect is based on the product of the corresponding correlation coefficients, as shown e.g. in (Boca et al., 2014):

**Eq. 4** $\quad E_2(x \to z \to y) = r_{xz} r_{yz \bullet x}$

where $r_{xz}$ denotes the correlation coefficient between "x" and "z", and $r_{yz \bullet x}$ denotes the partial correlation coefficient between "y" and "z" with the effect of "x" removed from both.

Throughout the present study, the mediation effect will be quantified by the product of correlation coefficients, as in Eq. 4, which has the appealing property of not depending on the scale (i.e. variance) of the variables. Later, this definition will be generalized to the complex-valued multivariate case.

## 4. The statistical test for mediation effect in the simple model

When testing the mediation effect, the null hypothesis of interest is:

**Eq. 5** $\quad H_0 : |E_2(x \to z \to y)| = |r_{xz} r_{yz \bullet x}| = 0$

In this paper, the method of choice for testing the mediation effect is the non-parametric randomization procedure proposed in (Taylor and MacKinnon, 2012). Let the triplets $(x_k, z_k, y_k)$, for $k = 1 \ldots N_R$, denote the sampled data. Then randomization of the mediator variable $z_k$ with respect to "k" will produce new randomized sampled data that satisfy the null hypothesis (Eq. 5), thus providing an estimator for the empirical probability distribution, from which the probability can be assessed.

This methodology can be immediately applied to the more general forms of mediation considered below, for complex valued multivariate case.

## 5. The general multivariate complex-valued mediation model

For the multivariate complex-valued case, let $\mathbf{X} \in \mathbb{C}^{p \times 1}$ denote the independent variable, $\mathbf{Z} \in \mathbb{C}^{q \times 1}$ the mediator variable, and $\mathbf{Y} \in \mathbb{C}^{r \times 1}$ the dependent variable, satisfying the relations:

**Eq. 6** $\quad \mathbf{Y} = \mathbf{BZ} + \mathbf{CX} + \boldsymbol{\varepsilon}_y$

**Eq. 7** $\quad \mathbf{Z} = \mathbf{AX} + \boldsymbol{\varepsilon}_z$

where $\mathbf{A} \in \mathbb{C}^{q \times p}$, $\mathbf{B} \in \mathbb{C}^{r \times q}$, and $\mathbf{C} \in \mathbb{C}^{r \times p}$ are regression coefficients, and $\boldsymbol{\varepsilon}_y \in \mathbb{C}^{r \times 1}$ and $\boldsymbol{\varepsilon}_z \in \mathbb{C}^{q \times 1}$ are noise.

In analogy with the approach of (Boca et al., 2014), as implemented in Eq. 4 and Eq. 5 above, we offer a more generalized definition which quantifies the mediation effect for the multivariate complex-valued as:

**Eq. 8** $\quad E_3(x \to z \to y) = RV_{xz} RV_{yz \bullet x}$

where the RV correlation coefficient of Escoufier (Escoufier, 1973; Robert and Escoufier, 1976) is used, which consists of a generalization of the common correlation coefficients that appear in Eq. 4 and Eq. 5.

The RV coefficient for the multivariate complex-valued case was introduced in (Pascual-Marqui et al., 2016) as:

**Eq. 9**
$$RV_{xz} = \frac{tr(\mathbf{S}_{zx}\mathbf{S}_{zy}^*)}{\sqrt{tr[(\mathbf{S}_{xx})^2]}\sqrt{tr[(\mathbf{S}_{zz})^2]}}$$

where $tr(\mathbf{M})$ denotes the trace of the matrix $\mathbf{M}$, the superscript "*" denotes conjugate-transpose, and where in general:

**Eq. 10**
$$\mathbf{S}_{uv} = \frac{1}{N_R}\sum_{k=1}^{N_R}\mathbf{U}_k\mathbf{V}_k^*$$

denotes the usual estimated Hermitian covariance for zero mean data, for sampled data indexed as $k=1...N_R$.

In analogy with the definition of the partial RV-coefficient for real-valued data given by (Vicari, 2000), we define the partial RV-coefficient for complex-valued data as:

**Eq. 11**
$$RV_{yz\bullet x} = \frac{tr\left[(\mathbf{S}_{yz}-\mathbf{S}_{yx}\mathbf{S}_{xx}^{-1}\mathbf{S}_{xz})(\mathbf{S}_{yz}-\mathbf{S}_{yx}\mathbf{S}_{xx}^{-1}\mathbf{S}_{xz})^*\right]}{\sqrt{tr\left[(\mathbf{S}_{yy}-\mathbf{S}_{yx}\mathbf{S}_{xx}^{-1}\mathbf{S}_{xy})^2\right]}\sqrt{tr\left[(\mathbf{S}_{zz}-\mathbf{S}_{zx}\mathbf{S}_{xx}^{-1}\mathbf{S}_{xz})^2\right]}}$$

Note that in the real-valued univariate case, where all vectors have only one component and are real-valued, the RV coefficients are equivalent to the square correlation coefficient and to the square partial correlation coefficient (see Eq. 4). Thus, for real-valued univariate "x", "y", and "z":

**Eq. 12** $\quad E_3(x \to z \to y) = \left[E_2(x \to z \to y)\right]^2$

**Eq. 13** $\quad RV_{xz} = r_{xz}^2$

**Eq. 14** $\quad RV_{yz\bullet x} = r_{yz\bullet x}^2$

When testing the mediation effect for the multivariate complex-valued case, the null hypothesis of interest is:

**Eq. 15** $\quad H_0 : E_3(x \to z \to y) = RV_{xz}RV_{yz\bullet x} = 0$

As previously explained, the method of choice for testing the mediation effect is the non-parametric randomization procedure proposed in (Taylor and MacKinnon, 2012). Let the triplets $(\mathbf{X}_k, \mathbf{Z}_k, \mathbf{Y}_k)$, for $k=1...N_R$, denote the sampled data. Then randomization of the mediator variable $\mathbf{Z}_k$ with respect to "k" will produce new randomized sampled data that satisfy the null hypothesis (Eq. 15), thus providing an estimator for the empirical probability distribution, from which the probability can be assessed.

## 6. A hypothetical mediation example: alpha "mediating" the transmission of beta

By way of illustration, consider the hypothetical example shown in Figure 2: widespread alpha oscillations (which can be measured in occipital regions $O\alpha$) mediate the "transmission" of local occipital beta oscillations ($O\beta$) to frontal regions ($F\beta$).

This example is intended to motivate the next section on cross-frequency mediation.

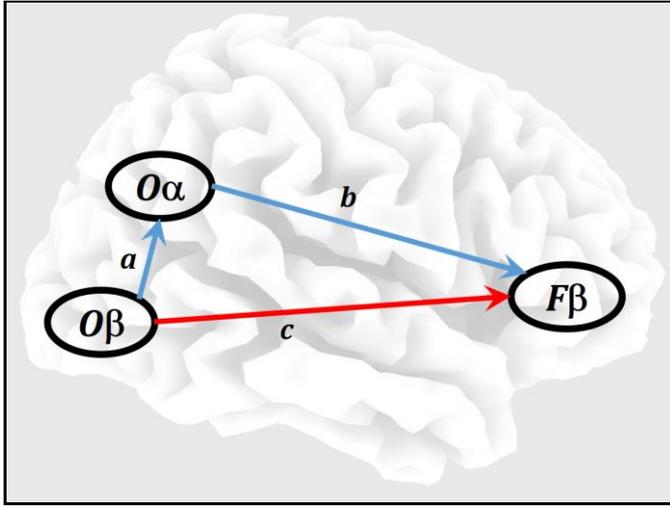

Figure 2: Hypothetical mediator model. Alpha (which is actually a widespread activity) is "mediating" the "transmission" of beta from occipital to frontal brain regions (the path with blue arrows).

## 7. The general cross-frequency mediation model

Let $x_k(t), z_k(t), y_k(t)$ denote three time series, for discrete time $t = 0\ldots(N_T - 1)$, with $k = 1\ldots N_R$ denoting the k-th time segment (i.e. epoch or time window). It will be assumed that these time series correspond to cortical electric neuronal activity.

Let:

**Eq. 16** $\quad x_k(\omega) = \sum_{t=0}^{N_T - 1} x_k(t) e^{-\iota 2\pi \omega t / N_T}$

denote the discrete Fourier transform (DFT) of "x", for discrete frequency $\omega = 0 \ldots N_T/2$, with $\iota = \sqrt{-1}$, with similar definitions for "z" and "y".

Let:

**Eq. 17** $\quad \Omega_1 = (\omega_{11}, \omega_{12}, \ldots, \omega_{1p})$

**Eq. 18** $\quad \Omega_2 = (\omega_{21}, \omega_{22}, \ldots, \omega_{2q})$

**Eq. 19** $\quad \Omega_3 = (\omega_{31}, \omega_{32}, \ldots, \omega_{3r})$

denote three frequency bands, each one defined by a collection of discrete frequencies. In this general notation, $\Omega_1$ consists of "p" discrete frequencies, $\Omega_2$ consists of "q" discrete frequencies, and $\Omega_3$ consists of "r" discrete frequencies.

Next, define the complex-valued vector:

**Eq. 20** $\quad \mathbf{X}_k(\Omega_1) = \begin{pmatrix} x_k(\omega_{11}) \\ x_k(\omega_{12}) \\ \ldots \\ x_k(\omega_{1p}) \end{pmatrix} \in \mathbb{C}^{p \times 1}$

with similar definitions for $\mathbf{Z}_k(\Omega_2) \in \mathbb{C}^{q \times 1}$ and $\mathbf{Y}_k(\Omega_3) \in \mathbb{C}^{r \times 1}$.

It is assumed throughout that all variables have zero mean.

Then the system:

Eq. 21 $\quad \mathbf{Y}_k(\Omega_3) = \mathbf{BZ}_k(\Omega_2) + \mathbf{CX}_k(\Omega_1)$

Eq. 22 $\quad \mathbf{Z}_k(\Omega_2) = \mathbf{AX}_k(\Omega_1)$

corresponds to a statistical mediation model, where $\mathbf{Z}_k(\Omega_2)$ is mediating the influence of $\mathbf{X}_k(\Omega_1)$ on $\mathbf{Y}_k(\Omega_3)$.

In descriptive words, this means that:
*the influence of "$\Omega_1$ activity at x", on "$\Omega_3$ activity at y", is partly mediated by "$\Omega_2$ activity at z".*

This mediator model (Eq. 21 and Eq. 22) implements a very general form of cross-frequency mediation, where one type of oscillatory activity at one place $\left[\mathbf{Z}_k(\Omega_2)\right]$, mediates a second type of oscillatory activity from a second place $\left[\mathbf{X}_k(\Omega_1)\right]$ to a third place $\left[\mathbf{Y}_k(\Omega_3)\right]$, with possibly change of oscillations (if $\Omega_3 \neq \Omega_1$).

Note that Eq. 22 is a simple linear model, which is very explicitly expressing cross-frequency coupling, from $\Omega_1$ oscillations to $\Omega_2$ oscillations. Eq. 21 is implementing an even more complicated form of cross-frequency coupling, from $\Omega_1$ and from $\Omega_2$ to $\Omega_3$.

In Eq. 21 and Eq. 22, the regression coefficients are complex valued, with $\mathbf{A} \in \mathbb{C}^{q \times p}$, $\mathbf{B} \in \mathbb{C}^{r \times q}$, and $\mathbf{C} \in \mathbb{C}^{r \times p}$.

The general cross-frequency mediation effect is $E_3(x \to z \to y)$ as defined above by Eq. 8, Eq. 9, and Eq. 11. The RV coefficients for Fourier transform coefficients was introduced in (Pascual-Marqui et al., 2016) as the "dual frequency RV-coupling coefficients". They correspond to a generalization of the "dual-frequency coherence" that was introduced by (Thomson, 1982; Haykin and Thomson, 1998). Note that the new definition introduced here for the partial RV coefficient in Eq. 11, applied to three different frequency bands in general, should be denoted as the "multi frequency partial RV-coupling coefficient".

### 8. Components of the general cross-frequency mediation effect: real and imaginary contributions, instantaneous and lagged mediation

As introduced in (Pascual-Marqui et al., 2016), note that the numerators in Eq. 9 and Eq. 11 have the general form:

Eq. 23 $\quad tr\left[\mathbf{MM}^*\right] = \sum_{i,j}\left|[\mathbf{M}]_{ij}\right|^2 = \sum_{i,j}\left\{\mathrm{Re}[\mathbf{M}]_{ij}\right\}^2 + \sum_{i,j}\left\{\mathrm{Im}[\mathbf{M}]_{ij}\right\}^2$

where $[\mathbf{M}]_{ij}$ denotes the complex-valued element (i,j) of the matrix $\mathbf{M}$, and $\mathrm{Re}[\bullet]$ and $\mathrm{Im}[\bullet]$ denote the real and imaginary parts of the argument.

This gives a natural decomposition of the "dual frequency RV-coupling coefficient" into contributions from the real and imaginary parts of the covariance. Thus, Eq. 9 can meaningfully be written as:

Eq. 24 $\quad RV_{xz} = RV_{xz}^{\mathrm{Re}} + RV_{xz}^{\mathrm{Im}}$

with the real and imaginary contributions defined as:

Eq. 25 $$RV_{xz}^{\text{Re}} = \frac{\sum_{i=1}^{q}\sum_{j=1}^{p}\left\{\text{Re}[\mathbf{S}_{zx}]_{ij}\right\}^2}{\sqrt{tr\left[(\mathbf{S}_{xx})^2\right]}\sqrt{tr\left[(\mathbf{S}_{zz})^2\right]}}$$

Eq. 26 $$RV_{xz}^{\text{Im}} = \frac{\sum_{i=1}^{q}\sum_{j=1}^{p}\left\{\text{Im}[\mathbf{S}_{zx}]_{ij}\right\}^2}{\sqrt{tr\left[(\mathbf{S}_{xx})^2\right]}\sqrt{tr\left[(\mathbf{S}_{zz})^2\right]}}$$

These components (Eq. 25 and Eq. 26) are equivalent, respectively, to the squares of the real and imaginary parts of the coherence for the case $p = q = 1$.

Similarly, "multi frequency partial RV-coupling coefficient" in Eq. 11 can be written as:

Eq. 27 $$RV_{yz \bullet x} = RV_{yz \bullet x}^{\text{Re}} + RV_{yz \bullet x}^{\text{Im}}$$

with:

Eq. 28 $$RV_{yz \bullet x}^{\text{Re}} = \frac{\sum_{i=1}^{r}\sum_{j=1}^{q}\left\{\text{Re}\left[\mathbf{S}_{yz} - \mathbf{S}_{yx}\mathbf{S}_{xx}^{-1}\mathbf{S}_{xz}\right]_{ij}\right\}^2}{\sqrt{tr\left[(\mathbf{S}_{yy} - \mathbf{S}_{yx}\mathbf{S}_{xx}^{-1}\mathbf{S}_{xy})^2\right]}\sqrt{tr\left[(\mathbf{S}_{zz} - \mathbf{S}_{zx}\mathbf{S}_{xx}^{-1}\mathbf{S}_{xz})^2\right]}}$$

Eq. 29 $$RV_{yz \bullet x}^{\text{Im}} = \frac{\sum_{i=1}^{r}\sum_{j=1}^{q}\left\{\text{Im}\left[\mathbf{S}_{yz} - \mathbf{S}_{yx}\mathbf{S}_{xx}^{-1}\mathbf{S}_{xz}\right]_{ij}\right\}^2}{\sqrt{tr\left[(\mathbf{S}_{yy} - \mathbf{S}_{yx}\mathbf{S}_{xx}^{-1}\mathbf{S}_{xy})^2\right]}\sqrt{tr\left[(\mathbf{S}_{zz} - \mathbf{S}_{zx}\mathbf{S}_{xx}^{-1}\mathbf{S}_{xz})^2\right]}}$$

These components (Eq. 28 and Eq. 29) are equivalent, respectively, to the squares of the real and imaginary parts of the partial coherence for the case $p = q = r = 1$.

Plugging these equations into Eq. 8 gives:

Eq. 30
$$\begin{aligned}E_3(x \to z \to y) &= RV_{xz} RV_{yz \bullet x} \\ &= \left(RV_{xz}^{\text{Re}} + RV_{xz}^{\text{Im}}\right)\left(RV_{yz \bullet x}^{\text{Re}} + RV_{yz \bullet x}^{\text{Im}}\right) \\ &= RV_{xz}^{\text{Re}} RV_{yz \bullet x}^{\text{Re}} + RV_{xz}^{\text{Re}} RV_{yz \bullet x}^{\text{Im}} + RV_{xz}^{\text{Im}} RV_{yz \bullet x}^{\text{Re}} + RV_{xz}^{\text{Im}} RV_{yz \bullet x}^{\text{Im}}\end{aligned}$$

These results are useful for treating the following problem. It is well known that signals of electric neuronal activity estimated from extracranial EEG / MEG recordings have low spatial resolution, see e.g. (Pascual-Marqui, 2007;Pascual-Marqui et al., 2011). This implies that the signals will be highly correlated at lag zero (i.e. instantaneously). It is of interest to take this into account, and to develop measures of connectivity that reflect physiology, without being confounded with this low resolution artifact.

One solution to this problem, as proposed by (Nolte et al., 2004), is to consider only the imaginary part of the coherence. In analogy with this approach, we can define the mediation effect for imaginary parts only as:

Eq. 31 $$E_4^{\text{Im}}(x \to z \to y) = RV_{xz}^{\text{Im}} RV_{yz \bullet x}^{\text{Im}}$$

Another solution to this problem, as proposed by (Pascual-Marqui, 2007;Pascual-Marqui et al., 2011), consists of expressing the total "connectivity" in terms of a non-instantaneous (i.e. lagged-physiological) component, and instantaneous components that include the low resolution artifact (involving all real part contributions). In analogy with this approach, see Equation 3.17 in (Pascual-Marqui et al., 2011), we can define the lagged physiological mediation effect as:

Eq. 32 $$E_5^{Lag}(x \to z \to y) = \frac{RV_{xz}^{Im} RV_{yz \bullet x}^{Im}}{1 - RV_{xz}^{Re} RV_{yz \bullet x}^{Re} - RV_{xz}^{Re} RV_{yz \bullet x}^{Im} - RV_{xz}^{Im} RV_{yz \bullet x}^{Re}}$$

## 9. The cross-frequency mediation model for multivariate time series

Note that this model (Eq. 21 and Eq. 22) can accommodate the case when the time series "x", "y", and "z" are multivariate, by simply stacking the multivariate vectors one under another in Eq. 20 for "x" (and similarly for the other time series). This will produce a mediation model of higher dimensions.

## 10. The statistical test for mediation effect in the general multivariate complex-valued model

When testing the mediation effect in general, the null hypotheses of interest are:

Eq. 33 $$\begin{cases} H_0 : E_3(x \to z \to y) = 0 \\ H_0 : E_4^{Im}(x \to z \to y) = 0 \\ H_0 : E_5^{Lag}(x \to z \to y) = 0 \end{cases}$$

As previously stated in this paper, the method of choice for testing the mediation effect is the non-parametric randomization procedure proposed in (Taylor and MacKinnon, 2012). Let the triplets $(x_k, z_k, y_k)$, for $k = 1 ... N_R$, denote the sampled data. Then randomization of the mediator variable $z_k$ with respect to "k" will produce new randomized sampled data that satisfy the null hypotheses (Eq. 33), thus providing an estimator for the empirical probability distribution, from which the probability can be assessed.

## 11. Discussion

In the early literature on EEG, each classical clinical frequency band had been assigned a specific type of function, but in reality, each band might be multi-functional. Furthermore, each band has been assigned a preferred cortical area, but in reality, each band might be widely distributed. Examples of the traditional view are: occipital alpha rhythm, motor mu-rhythm, and frontal midline theta. An excellent review is presented in (Groppe et al., 2013).

However, it is naive to think that the brain only uses a specific band in a specific cortical region for a specific purpose or function.

Extensive literature exists for brain activity time series studies that are based on the assumption of stationarity. This is due to the appealing properties of stationary processes, for which a form of "central limit theorem" exists, which states that the cross-spectrum contains all the relevant information on the process, and that the frequencies are independent. However, this approach has limited for a very long time the exploration of cross-frequency coupling.

The aim here was to develop a model that allows to study how the brain integrates the different oscillations across different cortical regions.

A very important recent publication demonstrates this type of mechanism in brain function. In (de Pesters et al., 2016), using ECoG recordings in humans and in one primate, it is shown that in a motor task, alpha activity decreases and gamma activity increases in task related areas, while alpha activity increases in even very distant non-task areas.

The tools developed here are intended to quantify this type of phenomena.

From an electrophysiological point of view, one may ask why do two different types of oscillations (e.g. alpha and gamma) suppress each other locally, at the same site? One possible explanation is that locally, the two oscillations must use the same resources, i.e. the same neuronal populations. And due to this limitation of resources, only one type of oscillation is dominant. Another essential question is why does alpha decrease at one location is usually followed by alpha increase at many other regions, see e.g. (Pfurtscheller and Lopes da Silva, 1999;Neuper and Pfurtscheller, 2001;Pfurtscheller, 2003;de Pesters et al., 2016)? One possible explanation is that this corresponds to a long-range lateral inhibition mechanism, which is likely a large-scale property of the cortex. In other words: a local alpha decrease corresponds to an increase in cortical excitability, which triggers a widespread lateral inhibition (decreased cortical excitability) implemented as an alpha increase at other cortical regions.